# Anomalous differential conductance of In-$Bi_2Te_3$ contact

R. Yang, G. Yu

*National Laboratory for Infrared Physics, Shanghai Institute of Technical Physics, Chinese Academy of Science, Shanghai 200083, People's Republic of China*

The differential conductance of In/$Bi_2Te_3$ contact under cryogenic temperature is investigated. Anomalous zero/finite bias differential conductance has been observed. The dependence of the observed anomalous differential conductance on magnetic field and temperature is studied. Data analysis indicates that the anomalous differential conductance is caused by the peculiar transport properties of the superconducting Indium/$Bi_2Te_3$ interface where Andreev reflection plays a key role.

# I. Introduction

The contact conductance between two materials seems to be very ordinary. However, it actually harbors many phenomena, such as the Schottky contact phenomenon and Ohmic contact phenomenon. One of the most interesting physical processes harbored in the contact conductance is the Andreev reflection. Andreev reflection happens in the interface between a superconductor and a normal material. In Andreev reflection, an incident electron whose energy is less than the superconductor's energy gap will be retroreflected as a hole, and a Cooper pair will be transmitted into the superconductor,[1-5] the process will also work with an incident hole (and retroreflected electron). This process is the origin of proximity effect[6] and determines the transport properties of the superconductor/normal material (S/N) contact. [1-6]

The transport features in the interface between a superconductor and normal material have been investigated in detail. Experimentalists observed characteristic conductance spectrum (differential conductance as a function of bias voltage) in various S/N contacts or S/N/S structures, such as Nb/InAs/Nb junction,[7] PbTe/In contact,[8] CdTe/In contact[9] and InGaAs/Nb contact. [4,10] The technique developed based on the Andreev reflection—Andreev spectroscopy, has been proved to be a powerful tool for the field of spintronics.[3,5] The spin polarization of various materials was characterized through Andreev spectroscopy.[3,5]

Due to the Andreev reflection, superconductor can serve as a 'barrier'.[4] Consequently, quasi-bound state (Andreev bound state, ABS) can exist in the well defined by the barriers where at least one barrier is superconductor. Such quasi-bound state was firstly investigated by de Gennes and Saint-James[11-13] and the conductance resonance caused by this kind of quasi-bound state is known as dGSJ resonance.[12] ABS could also exist in the interface between a superconductor and vacuum.[13,14] The ABS on the surface of a superconductor can give rise to characteristic conductance spectrum.[13,14] A zero bias conductance peak (ZBCP) in the Andreev spectroscopy of nonconventional superconductors was reported in previous experimental or theoretical investigations.[13,14] This peculiar conductance spectrum has been employed to verify the existence of ABS and characterize the properties of nonconventional superconductors.[13,14] ABS can also be employed to realize ABS qubit for quantum computation.[15]

Additionally, in recent years, experiments and theories have pushed the frontier of the field of Andreev physics into a fantastic new world.

The discovery of graphene has brought new vibrancy into the field of condensed matter physics. The weird quasi-relativistic particles existing in this material harbor numerous new physics.[16-20] The graphene/Superconductor hybrid structure has been revealed to exhibit many interesting transport properties.[20-22]

Moreover, the investigation of ABS also led to the discovery of Majorana physics. Majorana Fermion is an amazing quasiparticle which has great potential in the field of fault-tolerant topological quantum computation due to its non-Abelian statistics.[23,24] In the vortex cores of p+ip superconductors, Majorana Fermions would emerge.[25-28] On the surface of some unconventional superconductors with special pairing symmetry, the spin-orbit interaction will also lead to the emergence of bound state of Majorana Fermion.[14] Comparing with other quantum computation schemes which are struggling in the mud to protect their fragile quantum coherence, topological quantum computation is relatively robust against decoherence.[24,29-31] The existence of Majorana Fermion can leave fingerprint signatures in the conductance spectrum.[23] Recently, a new kind of material—topological insulator (TI) where strong spin-orbit interaction plays a key role has been discovered.[32-39] In this new perspective, some conventional narrow-gap semiconductors such as thermoelectric material $Bi_2Te_3$ and $Bi_2Se_3$ have been verified to be TI.[35,37] With the discovery of topological insulator, the investigation of transport processes in S/N structure has been extended into S/TI system.[23,24,40-42] Many features in the transport process of such structures were revealed.[23,24,40-42] Especially, the aforementioned mechanism giving rise to the emergence of Majorana Fermions has been proved to also exist in the interface between a TI and a superconductor.[23,24,40] Therefore, S/TI structure serves as a possible platform for the realization of non-Abelian physics and the fault-tolerant topological quantum computation.[23,24]

Anyway, the differential conductance spectrum of the superconductor/$Bi_2Te_3$ contact is a problem worth investigating. In this paper, we investigated the differential conductance of In-$Bi_2Te_3$ contact. Anomalous differential conductance caused by superconductor Indium/$Bi_2Te_3$ junction has been observed. The dependence of the observed anomalous differential conductance on external magnetic field and temperature is studied. Possible explanations for the observed anomalous differential conductance are provided.

**II. Sample fabrication and measuring system**

Our samples are p-type $Bi_2Te_3$ single crystal doped with Sn (with nominal Sn concentration 0.5%). The shape of our samples are 0.5 mm thick films, with the surfaces perpendicular to the C axis ([0001] axis) of the hexagonal structure of $Bi_2Te_3$. XRD and transport measurements are employed to characterize our samples. Indium is used to form contacts. The magnetoresistance is measured under Van der Pauw configuration and the magnetic field is applied perpendicular to the film and then tilted. All measurements are performed in an Oxford Instruments $^4$He cryogenic system with temperatures ranging from 1.3 to 4.52 K. A package of Keithley sourcemeters is used to provide current and to measure magnetoresistance.

**III. Experimental results and discussions**

Fig. 1 presents the XRD of our sample, we can distinguish many characteristic XRD peaks of $Bi_2Te_3$ crystal. A feature of the XRD is that the last index is multiples of five. It's caused by the characteristic systematic absence of $Bi_2Te_3$ crystal when X-ray is incident towards [001] crystal plane.

Fig. 2 presents the magnetoresistance of our sample, the magnetoresistance is measured under van der Pauw configuration. In Fig. 2 (a), we can see obvious SdH oscillations. In Fig. 2 (b), the dependence of $\rho_{xy}$ on magnetic field indicates that our sample is p-type. The emergence of SdH oscillations and quantum Hall plateaus is also consistent with previous studies regarding the magnetotransport properties of Bi-based semiconductor.[43-48] For example, Kulbachinskii *et al.* observed SdH oscillations and quantum Hall effect in p-type $Bi_2Te_3$:Sn crystals with Sn doping same as our sample.[46,48] James, *et al.* observed SdH oscillations in topological insulator $Bi_2Se_3$.[47,49] Taskin, *et al.* observed oscillations in topological insulator $Bi_xSb_{1-x}$.[44] In the context of topological insulator, a few groups have related some features in the magnetoresistance to the surface state of topological insulator,[44,47,49] though the samples in these experiments are not real bulk insulator. At high temperatures, strong magnetic fields and large bias voltages, the contact between Indium and $Bi_2Te_3$ is Ohmic contact. It is consistent with the fact that p-type conduction material would help facilitate the formation of Ohmic contact when contacting with metal.

For temperatures range from 1.3 K to ~4 K, we observe zero and finite bias anomalies of the differential conductance in two $Bi_2Te_3$ samples (sample 1 and sample 2) when the current is applied through the leads in which at least one lead is made of Indium. In sample 1, all the leads are made of Indium. In sample 2, some of the leads are made of Bismuth and some of the leads are made of Indium.

In order to nail down the origin of these anomalies, we also investigate the differential conductance in different setups. When all the leads in sample 2 are made of Bismuth, there are not any anomalies; when at least one lead is made of Indium, the zero/finite bias anomalies emerge. This control experiment indicates that this kind of anomaly is related with Indium.

In order to exclude other sources of anomalous differential conductance, we also investigate the differential conductance when the sample is replaced with a Cu plate. No anomaly is found in this setup. This observation excludes the possibility that the zero/finite bias anomalies observed in $Bi_2Te_3$ is caused by Indium alone, the contact between Indium and the wire, or the contact between Indium and the Cu base. Taking both of the aforementioned experiments into consideration, it's safe to conclude that the zero/finite bias anomalies observed in the $Bi_2Te_3$ sample originate from the contact between Indium and $Bi_2Te_3$.

The differential conductance spectrum of sample 1 and 2 are qualitatively the same. We take the data of sample 2 as an example, see Fig.3. In Fig.3, we can see a zero bias peak and two finite bias dips. Fig. 3 (a) presents the differential conductance as a function of bias voltage at different temperatures. Fig. 3 (b) presents the differential conductance at different external magnetic fields. We can see that, as the magnetic field and temperature increase, the anomalous zero bias differential conductance peak and finite bias differential conductance dip gradually vanish and finally disappear above certain magnetic field and temperature. The temperature (~4 K) and magnetic field (~40 mT) above which the anomalies disappear are consistent with that of superconducting Indium (3.41 K and 28.2 mT respectively). It indicates that these anomalies are related with superconducting Induim/Bi$_2$Te$_3$ contact. Similar results have also been observed in p-PbTe/In contact.[8]

For S/N contact where non-tunneling process dominates the transport, there are two configurations. One configuration is the ballistic point-contact configuration where the contact area is less than the Sharvin limit (the diameter of the contact area is less than the mean free path).[1, 50, 5] The other configuration is the diffusive S/N junction configuration where the transport is diffusive.[51] In the latter configuration, for the quantum interference effect caused by multiple scattering along with the Andreev reflection, two regimes exist.[51] One is the enhanced weak localization regime where the barrier is low (the interface is transparent, the transmission probability is near to unity).[51, 1, 52] In this regime, there is a differential conductance dip in the vicinity of zero bias.[51, 1, 52] The other regime is the reflectionless tunneling regime where the barrier is relatively high (the interface is not very transparent).[51, 1] In this regime, there is a differential conductance peak around the zero bias.[51, 1] If the transparency is further reduced towards zero, the peak will gradually flatten until vanish.[53, 54] The conductivity peak around zero bias arises from the constructive quantum interference of quasiparticles coherently backscattered towards the interface barrier in the diffusive region.[10] For S/N contact where tunneling process dominates, the differential conductance depends on the properties of interface. For most S/N contact, a zero bias dip in the differential conductance spectrum will develop. However, previous investigations revealed that the ABS on the surface of nonconventional superconductor (like d-wave paring superconductor) would lead to the emergence of a zero bias peak.[23] Moreover, Linder, *et al.* investigated the transport properties of the interface between a topological insulator and spin singlet superconductor (Indium is just a spin singlet/s-wave superconductor), their research reveals that ABS just as the ABS on the surface of a d-wave superconductor also exist in the interface, therefore, a peculiar ZBCP will exist.[23] Speaking of our experiment, the observed zero bias peak in our experiment can be explained with the reflectionless tunneling [51, 1] mechanism or the characteristic differential conductance spectrum of TI/s-wave Sc junction predicted by Ref. 23.

Similar conductance minimas ('valleys') observed in our experiments (see, Fig. 3) at finite bias voltages have also been observed in other experiments.[3, 55, 5] They can be explained by a phase-transition of the contact from the S/N state to the N/N state induced by a combination of Joule heating and the magnetic field generated by the current.[3, 55] In the S/N state at low voltage, the current is larger in magnitude than the corresponding current in the N/N state due to the Andreev reflection process. As the magnitude of the current increases, the dissipated power and the magnetic field increases, eventually inducing the transition from the S/N to the N/N state. In the transition regions, the slope of the current is smaller than that in the other regions, leading to the minima in the conductance curve.[3]

In Fig. 3 (b), we can see that the bias voltage at which the phase transition happens decreases as the environment magnetic field increases. The explanation for this phenomenon is simple. According to Ref. 3, phase transition happens when the magnetic field felt by Indium (which includes magnetic field caused by the bias current as well as the environment magnetic field) surpasses the critical magnetic field of Indium. Therefore, the increase of environment magnetic field would lead to the decrease of the bias voltage at which the phase transition happens. The magnitude of magnetic field where the finite bias anomaly disappears is consistent with the critical magnetic field of superconducting Indium (28.2 mT).

**IV. Conclusions**

In conclusion, the observed anomalous differential conductance can be explained with the physics revealed in the S/N contact. The anomalous finite bias differential conductance dips can be attributed to the phase transition of the contact from the S/N state to the N/N state induced by a combination of Joule heating and the magnetic field generated by the current. One plausible explanation for the anomalous ZBCP is the reflectionless tunneling process, but the possibility that this ZBCP is just the characteristic ZBCP of TI/s-wave Sc junction predicted by Ref. 23 cannot be excluded. Therefore, this weird ZBCP deserves further study.

**Figure captions**

Figure 1. XRD of $Bi_2Te_3$:Sn single crystal, inset is an illustration of its crystal structure[35].

Figure 2. Magnetoresistance of the Bi2Te3:Sn sample. (a)Transverse magnetoresistance $\rho_{xy}$ as a function of magnetic field. (b)Longitudinal magnetoresistance $\rho_{xx}$ as a function of magnetic field.

Figure 3. Differential conductance as a function of bias voltage under different temperatures (a) and magnetic fields (b). (c) A comparison between I~V curves obtained at 1.32 K and 4.52 K

Figure 1

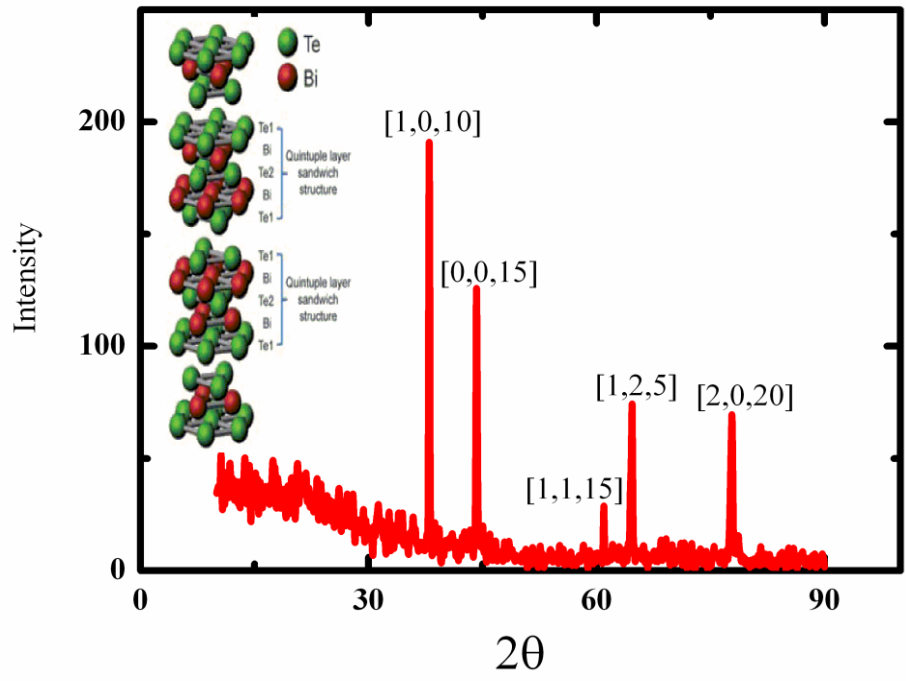

Figure 2 (a)

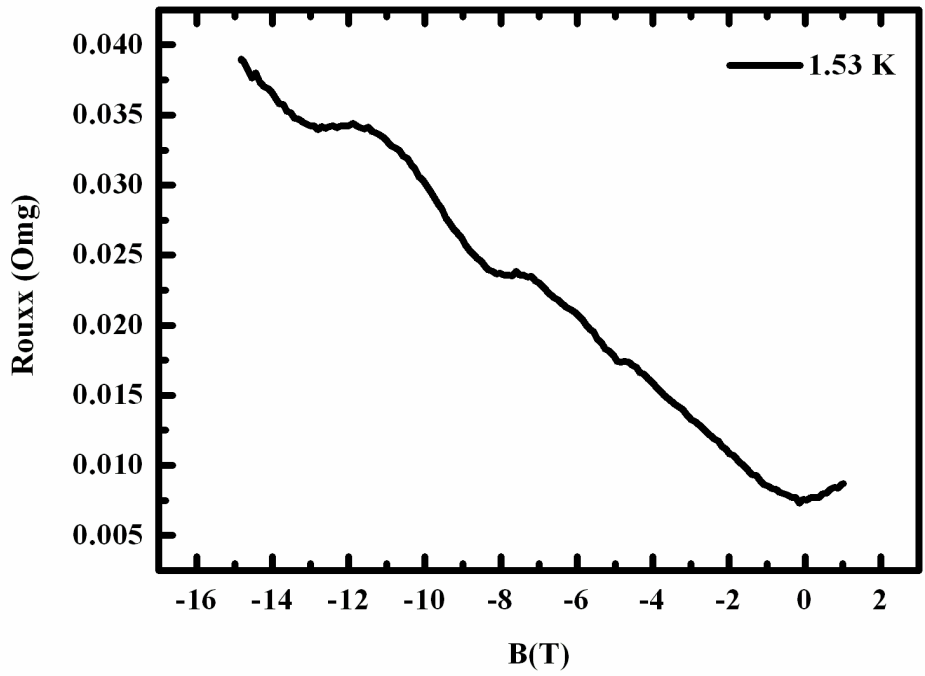

(b)

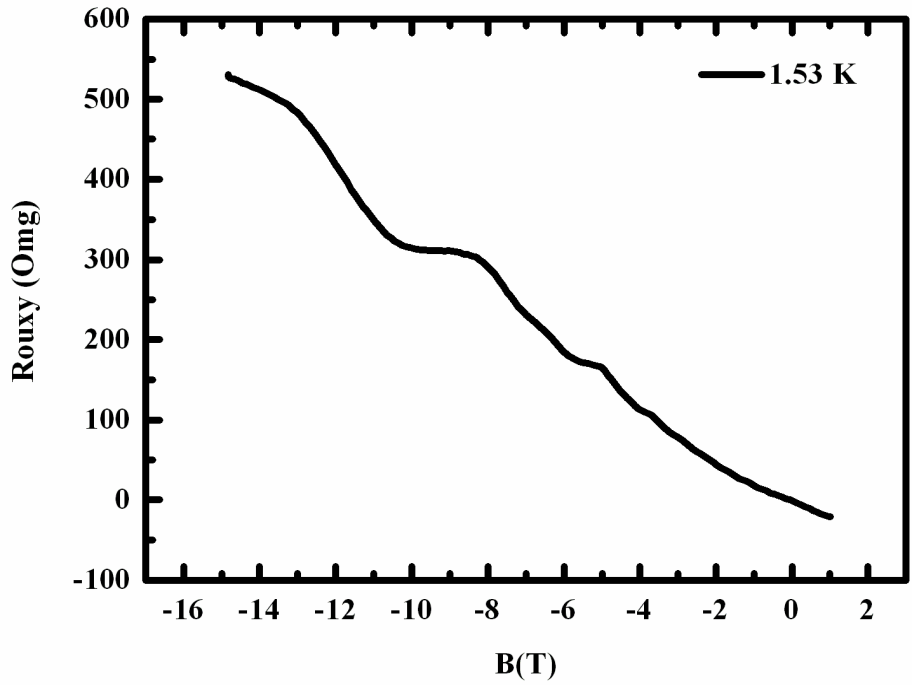

Figure 3 (a)

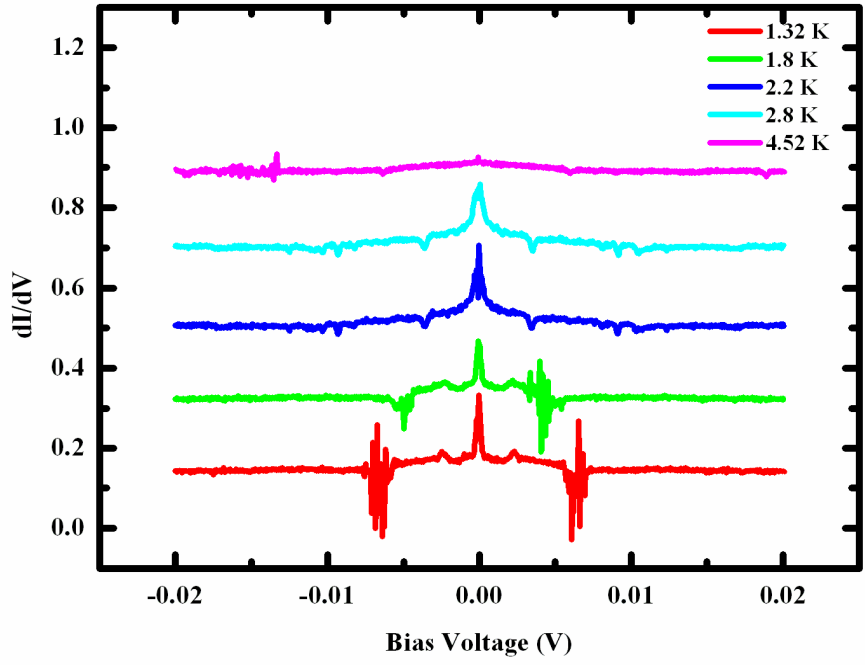

(b)

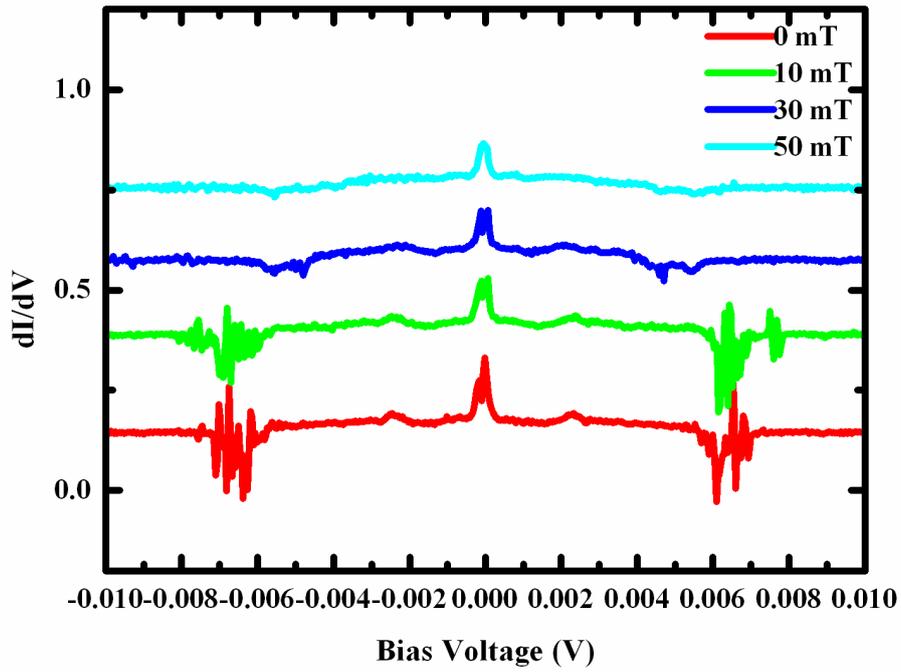

(c)

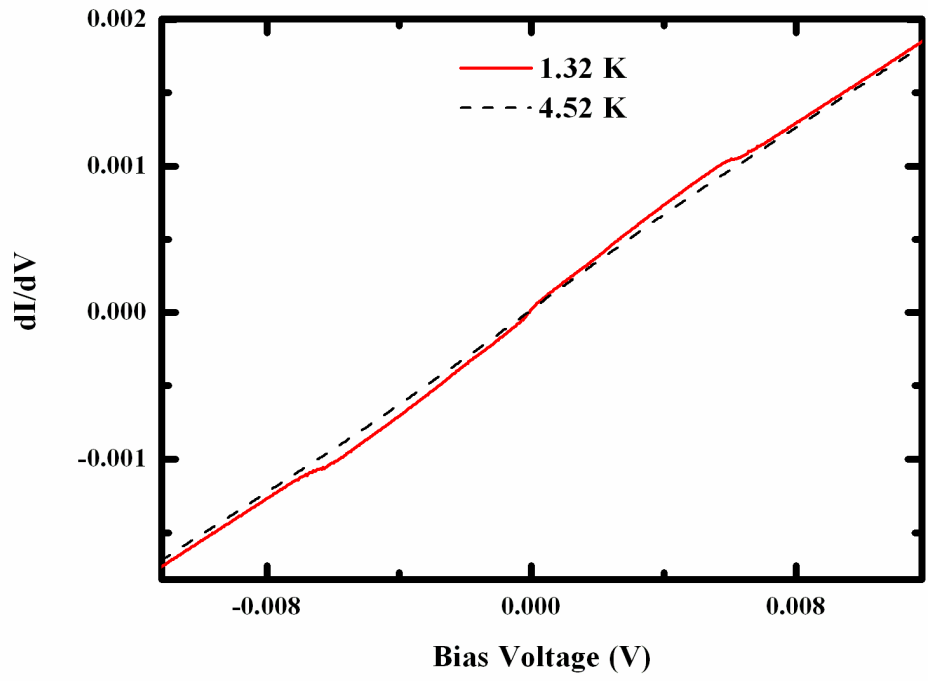